\documentclass[twocolumn,showpacs,amsmath,amssymb,prl,superscriptaddress]{revtex4}

\usepackage{amsmath}
\usepackage{graphicx}
\usepackage{amssymb}
\usepackage[T1]{fontenc}
\usepackage[utf8]{inputenc}
\usepackage{hyperref}
\usepackage{color}
\usepackage{float}

\newcommand{\ie}{\mathrm i}

\definecolor{meinmagenta}{rgb}{0.8, 0.16, 0.56}
\definecolor{meinorange}{rgb}{1, 0.27, 0}
\definecolor{meingruen}{rgb}{0, 0.55, 0.27}
\definecolor{meinhellblau}{rgb}{0, 0.75 ,1}
\definecolor{meindunkelblau}{rgb}{0, 0, 0.55}

\begin{document}


\title{On the stability and quality of power grids subjected to intermittent feed-in}



\author{Katrin Schmietendorf}
\affiliation{Institute for Physics, ForWind, Carl-von-Ossietzky-Universität, 26111 Oldenburg }
\author{Joachim Peinke}
\affiliation{Institute for Physics, ForWind, Carl-von-Ossietzky-Universität, 26111 Oldenburg }
\author{Oliver Kamps}
\affiliation{Center for Nonlinear Science, Westf\"alische Wilhelms-Universit\"at, 48149 M\"unster}

\date{\today}

%
%

\begin{abstract}
Feed-in fluctuations induced by renewables are one of the key challenges to the stability and quality of electrical power grids. 
In particular short-term fluctuations disturb the system on a time scale, on which  
load balancing does not operate yet and the system is intrinsically governed by self-organized synchronization. 
Wind and solar power are known to be strongly non-Gaussian with intermittent increment statistics in these time scales.
We investigate the impact of short-term wind fluctuations on the basis of a Kuramoto-like power grid model
considering stability in terms
of desynchronization and frequency and voltage quality aspects.
We compare intermittent feed-in with a realistic power spectrum, 
correlated Gaussian noise of the same spectrum, and Gaussian white noise. 
We found out that the likelihood of severe outages is mainly determined by the temporal correlation
of the feed-in. 
The intermittent nature of wind power is transferred into frequency and voltage fluctuations. 
This establishes a novel type of fluctuations with
severe consequences on frequency and voltage quality, which are beyond engineering status of knowledge.

\end{abstract}

\maketitle

\textit{Introduction.}--
Reliable energy supply is a basic requirement for modern society. In the course of the energy transition, the electrical power system is
undergoing substantial changes: Conventional power plants, whose power output is constant and adjustable to
the current demand, are progressively being replaced by fluctuating renewables like wind and solar.\\ 
\indent Wind power is one of the main renewable energy sources and intended to constitute a considerable part of
future energy composition.
It features power fluctuations ranging from seasonal to second and sub-second scale. 
These short-term fluctuations are particularly problematic, since they occur on a time-scale, on which load balancing
mechanisms do not operate yet and the power system is mainly governed by inertia.
As driven by athmospheric turbulence, wind power is strongly non-Gaussian with intermittent increment statistics and a $S(f)\sim f^{-5/3}$ power 
spectrum \cite{anvari2016njp,milan2013prl,apt2007journalOfPowerSources,calif2014nonlinearProcessesInGeophysics}.
Due to spatial correlations, the intermittency is not removed even on country-wide scales \cite{kamps2014windEnergy}.
If conventional supply shall be substituted by renewables, the influence 
of these specific fluctuations on grid stability and quality has to be understood.\\ 
\indent To capture the main characteristics of real wind feed-in,
we generate intermittent time series on the basis of a Langevin-type model and impose a realistic power spectrum. 
For comparison, we use correlated Gaussian noise of the same spectrum and Gaussian white noise.
The stochastic feed-in is implemented into a Kuramoto-like (KM) power grid model, capturing frequency and 
voltage dynamics in the order of seconds.
The KM approach is based on electrical engineering standards \cite{kundur1994book,machowski2008book} and has been seized by 
nonlinear dynamics 
and control theory to investigate basic mechanisms of power system dynamics and stability-topology interplay in recent
years (e.\,g. \cite{filatrella2008epjb,doerfler2010americancontrolconf,witthaut2012njp,menck2014natcomm,motter2013nature}).
The impact of grid decentralization, the second key issue of future power grids, was considered within this 
framework in \cite{rohden2012prl}.
With this study, we initiate to investigate the impacts of stochastic feed-in with realistic properties: temporal correlation,
realistic power spectrum and intermittent increment statistics.
We consider two aspects of practical relevance: the likelihood
of noise-induced desynchronization given by the average escape time, and frequency and voltage quality.

\begin{figure}[b]
 \includegraphics[scale=0.6]{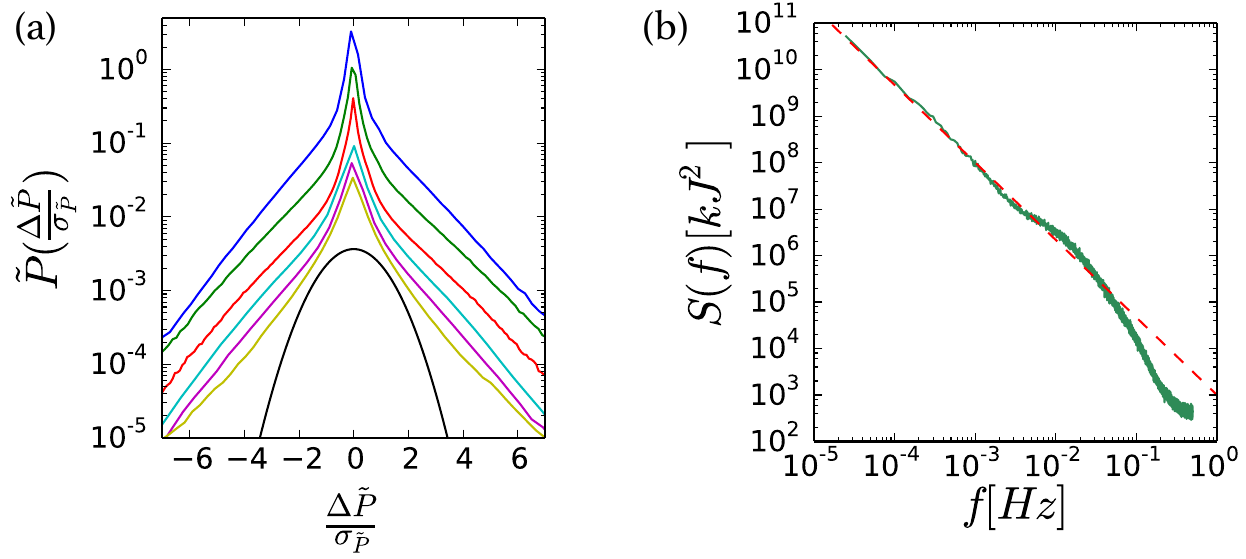}
\caption{Power output of a wind farm with 12 power plants \cite{milan2013prl}: (a) Pdfs of power increments $\Delta \tilde P=\tilde P(t)-\tilde P(t-\tau)$ for $\tau$=1s, 5s, 10s, 
50s, 100s, Gaussian distribution (from top 
to bottom, vertically shifted for clarity). (b) Power spectrum $S(f)$, red line indicates $-\frac 53$ decay.}
\label{fig:real_data}
\end{figure}

\begin{figure}[t]
 \includegraphics[scale=0.43]{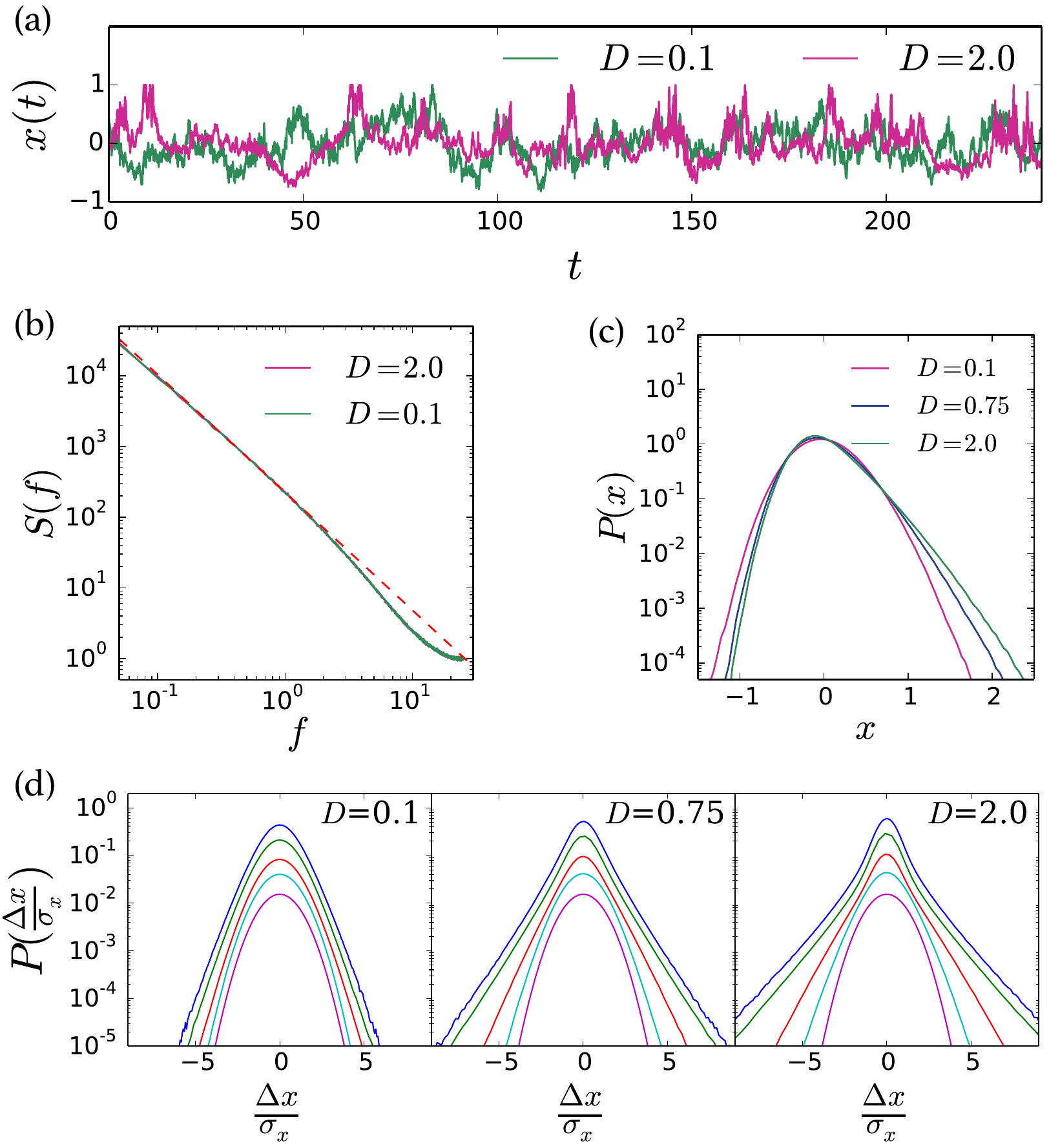}\\
\caption{Intermittent noise with $\gamma=1.0$, $g=0.5$ and $x_0=2.0$, normalized by 4.0.
For all $D\in[0.1,2.0]$, $\langle x\rangle=0.0$ and $\sigma_x$=0.328.
(a) Time series $x(t)$ (restricted to $[-1.0,1.0]$, see below).
(b) Power spectrum $S(f)$, the red dotted line indicates $\frac 53$ decay).
(c) Pdfs $P(x)$.
(d) Normalized pdfs of the increments $\Delta x=x(t)-x(t-\tau)$ for time lags 
$\tau=0.02, 0.2, 2.0, 20.0$, Gaussian distribution (top to bottom).}
\label{fig:syn_data}
\end{figure}

\indent \textit{The power grid model.}--
In power system dynamics, a synchronized state with constant frequencies, voltages and stationary power transfer
is the desired mode of operation.
Within the KM-like approach, the power grid is represented by a network of $N$ synchronous generators and motors transforming 
mechanical power into electrical power, or
vice versa. The coupled dynamics of the phase angles $\{\delta_i\}$ and magnitudes $\{E_i\}$ 
of the complex nodal voltages $\{\boldsymbol E_i=E_i\mathrm{e}^{\ie \delta_i}\}_{i\in\{1,..,N\}}$ are given by \cite{schmietendorf2014epjsti}:
\begin{eqnarray}\label{eq:networkEqs}
 \ddot\delta_i&=&-\gamma_i\dot\delta_i+P_{\mathrm m,i}-\sum\limits_{j=1}^N B_{ij}E_iE_j\sin\delta_{ij},\\
 \alpha_i\dot E_i&=& C_i-E_i+\chi_i\sum\limits_{i=1}^N B_{ij}E_j\cos\delta_{ij},
\end{eqnarray}
with $\omega_i=\dot\delta_i$ being the individual frequency of the $i$-th oscillator. $P_{\mathrm m,i}$ denotes the mechanical input or 
output power and $P_{\mathrm e,ij}=B_{ij}E_iE_j\sin\delta_{ij}$ is the electrical real power transferred between machines $i$ and $j$.
The susceptance matrix $\{B_{ij}\}_{i,j\in\{1,..,N\}}$ represents the network topology in case of a lossless high transmission grid.
$\gamma_i$, $\alpha_i$, $C_i$, $\Gamma_i$ and $\chi_i$ comprise machine and line parameters.\\
\indent\textit{Network topology.}--
\begin{figure}[t]
 \includegraphics[scale=0.36]{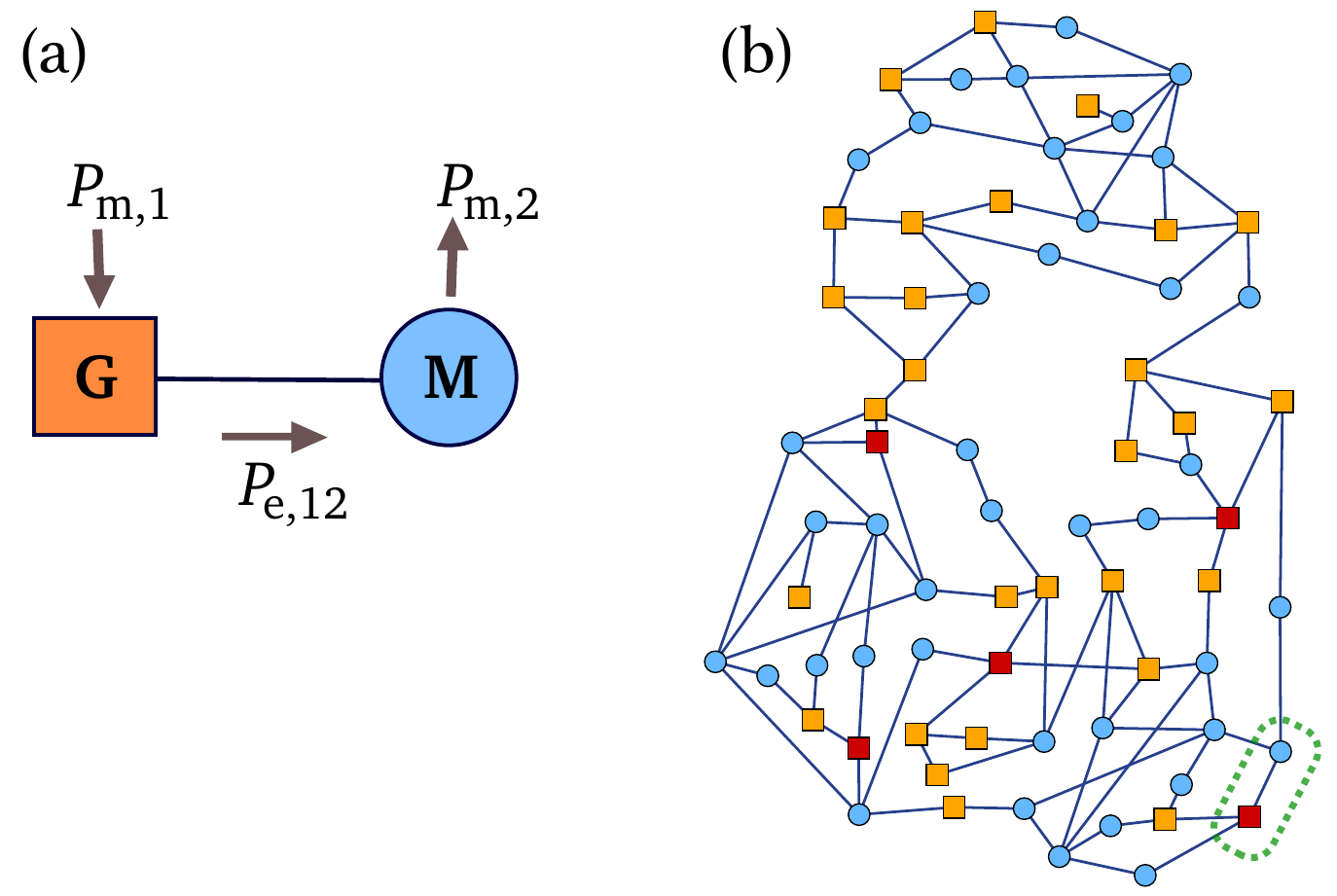}
\caption{(a) Two-machine system: generator G with input $P_{\mathrm m,1}$ transferring electrical power $P_{\mathrm e,12}$ to a motor
M with output $P_{\mathrm m,2}$. (b) IEEE topology: orange/red squares denote generators, 
blue circles motors. The red squares and green dashed box indicate the fluctuating generators and the selected G-M component
for the investigations on voltage and frequency quality.}
\label{fig:topologies}
\end{figure}
We first consider a two-machine (G-M) system as the basic componenent of 
any power network.
The complex network topology is based on an IEEE test system with 33 generators, 40 consumers and 118 links 
(see Fig.\,\ref{fig:topologies})
\footnote{We extracted the adjacency matrix and the topological 
location of the generators and consumers from the full dataset, see \cite{grigg1999ieeeTransactions}.}.
The machine parameters are set equal: $\gamma_i=0.2$, $\alpha_i=2.0$, 
$C_i=0.993$,  $\chi_i=0.1$. $P_{\text m,i}$ and $B_{ij}$ are specified below.\\
\indent\textit{Implementing noise.}--
The increment probability density functions (pdf) of real wind power data significantly 
deviate from Gaussianity and its power spectrum $S(f)$ displays $\frac 53$-decay with some discrepancy in the high frequency range
(see Fig.\,\ref{fig:real_data}).
Based on this, we consider three types of synthetique feed-in noise. For the most realistic scenario, 
we generate intermittent power time series by use of the Langevin-type model \cite{kamps}:
\begin{equation}\label{eq:noise_y}
 \dot y=-\gamma y+\Gamma(t),\quad \dot x=x\left( g-\frac{x}{x_0}\right)+\sqrt{Dx^2}y.
 \end{equation}
In a second step, the spectrum is modified in order that it essentially 
resembles the real data power spectrum.
Fig.\,\ref{fig:syn_data} shows the characteristics of the resulting power time series. 
The parameter $D$ controls the intermittence strength:
From $D=0.1$ (weakly intermittent, nearly Gaussian) to $D=2.0$ (strongly intermittent),
the time series become burstier with more heavy-tailed increment pdfs.  
Next, we drop the intermittency feature and use Gaussian noise of the same power
spectrum and standard deviation. By enforcing the spectrum we induce temporal correlations. This type of noise is 
referred as \textit{Gaussian53} in the following.
Thirdly, we put aside the applicational requirements and consider Gaussian white noise.\\ 
\indent We implement feed-in fluctuations into Eq. (\ref{eq:networkEqs}) by replacing the generator's constant power input  
by $\tilde P(t)=P_{\text m}+px(t)$.
The penetration $p$ allows to tune the percentage of fluctuating input.
Since $\langle x(t)\rangle=0$, $\langle\tilde P(t)\rangle=P_{\text m}$ and power-load balance is maintained on long-time average. 
We further integrate two properties of real  
wind power injection into the feed-in time series. First, the generator is not allowed to act as a motor, 
i.\,e. $P(t)>0$ has to be assured. Secondly, there is a cutoff for the maximum (rated) power fed into the grid. 
Therefore, we restrict $x(t)$ to $[-1.0,1.0]$ by setting $x(t)=-1.0$ $\forall$ $x(t)<-1.0$ and 
$x(t)=1.0$ $\forall$ $x(t)>1.0$. 
This additionally truncates some of the extreme events in the strongly intermittent power time series.
The mean and standard deviation is nonetheless nearly unaffected by this.


\indent\textit{Stability assessment.}--
In electrical engineering, power system stability is customarily defined as the system's ability to return to stationary operation 
after a specific disturbance \cite{hill2004ieee_transactions}. In other
words, it depends on whether the system returns to a synchronized state with constant phase angles $\delta_i$, 
frequencies $\omega_i$, and voltages $E_i$, i.\,e. to a fixed point. As stability measures for 
deterministic KM-like power systems, \textit{basin stability} \cite{menck2013natphys} and 
\textit{survivability} \cite{hellmann2016scientificReports} have 
been established. The former quantifies the
likelihood of returning to stationary operation after a random shift in phase space, the latter additionally monitors tolerance bound violations.
\indent Stochastic feed-in causes parameter fluctuations, which permanently modify the fixed point's basin
of attraction.
Hence, the system variables fluctuate around their stationary values under corresponding constant input. 
The system may even be kicked out of the basin of attraction, which implies desynchronization and therefore can be 
interpreted as a severe system outage.
The \textit{average escape time} provides a stability measure related to the probability of such incidents.
From the engineering viewpoint, frequency and voltage quality play an important role in power system assessment. 
Not only violations of tolerance bounds are relevant,
but also deviations from nominal values in general and their corresponding gradients, 
since they stress elctronic devices and may cause flickering  
\cite{entsoeHandbookPolicy1,ibrahim2011energyProcedia}.
These informations are revealed by frequency and voltage pdfs and the corresponding increment pdfs.

\begin{figure}[t]
\includegraphics[scale=0.72]{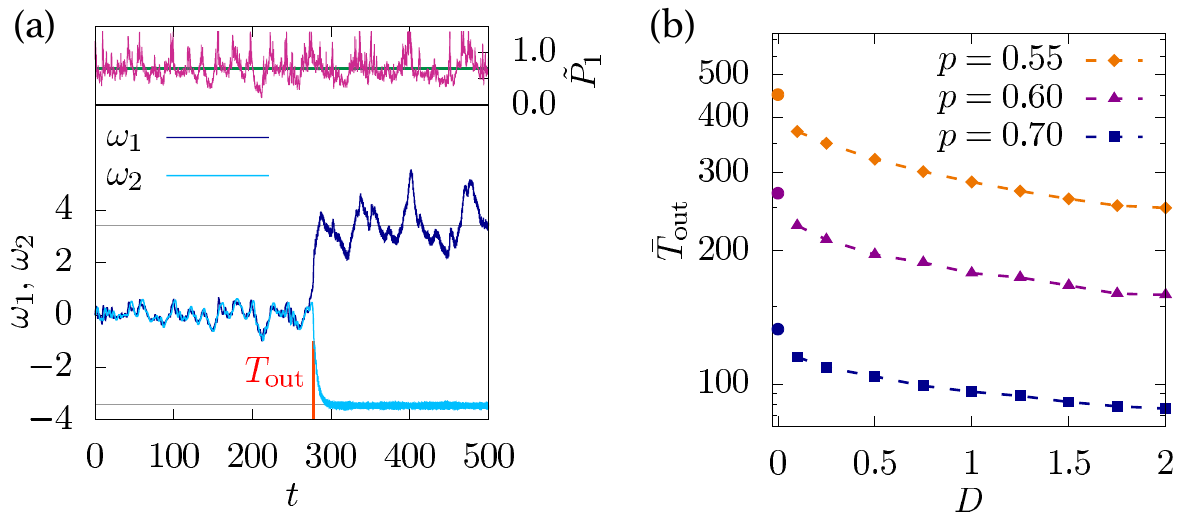}
\caption{Escape scenario in the two-machine system. (a) Generator input $\tilde P_1(t)=P_{\text m,1}+px(t)$ ($D$=2.0 
and $p$=0.7),
generator frequency $\omega_1$ and motor frequency
$\omega_2$, the grey lines indicate the location of the stable limit cycle. Escape at 
$T_\text{out}=283$. (b) $\bar T_\text{out}$ as a function of $D$ for $p=0.55$, 0.6, and 0.7, calculated from 
$M$=10000 trajectories. Circles denote $\bar T_\text{out}$ for Gaussian53 noise.}
\label{fig:T_out_D_twoMachine}
\end{figure}
\begin{figure}[t]
\includegraphics[scale=0.72]{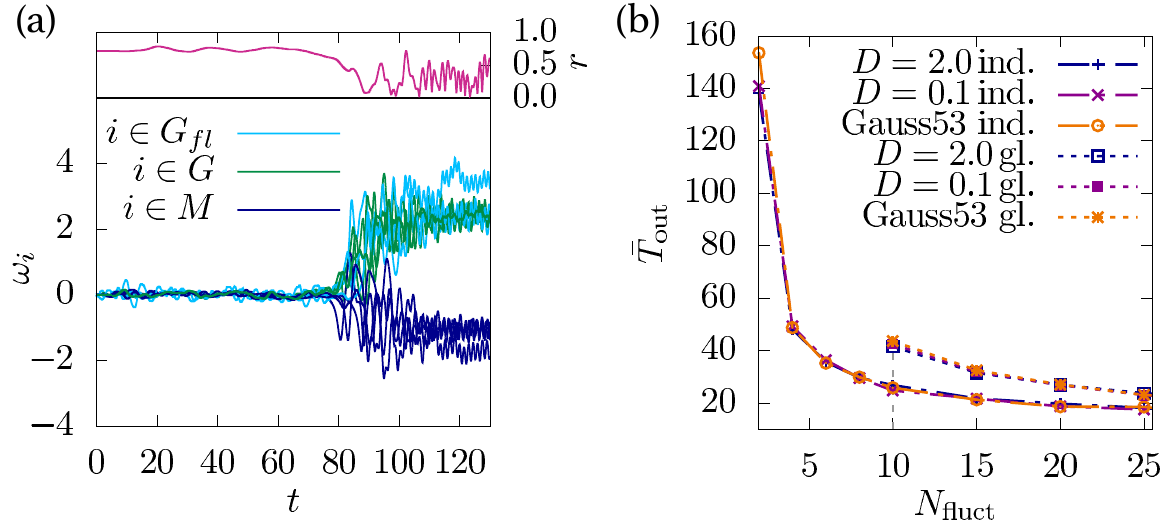}
\caption{Escape scenario in the IEEE grid: (a) Phase coherence $r$ 
 and frequencies $\omega_i$ for the first twelve machines, that fall out. Line colours indicate the machine type. 
 (b) $\bar T_\text{out}$ as a function of $N_\text{fl}$. Strongly intermittent ($D=2.0$), weakly intermittent ($D=0.1$) and 
 Gaussian53 noise, each individual and global, calculated from $M$=25000 trajectories. 
For global noise and $N_\text{fl}<10$ stable configurations occur, which survive simulation intervals of $T>30000$.}
\label{fig:T_out_Nfl}
\end{figure}

\textit{Two-machine system.}--
The G-M system (Fig.\,\ref{fig:topologies}) is operated in a coexistence region of a stable fixed point and a stable limit 
cycle \cite{schmietendorf2014epjsti} for $P_{\text m,1}$=$-P_{\text m,2}$=0.7, $B_{12}=1.0$ and $B_{11}=B_{22}=-0.8$.
For constant feed-in the system runs in its fixed point ($\omega^*_{1}=\omega^*_{2}=0$ 
in a reference frame rotating with nominal frequency). In the following, the terms \textit{fixed point}
and \textit{stationary values} refer to the corresponding system under constant power input ($p=$0).
Fig.\,\ref{fig:T_out_D_twoMachine} shows an example escape scenario.
First, generator and motor frequency fluctuate
around their fixed point values. At $T_\text{out}$ the system leaves its basin of attraction and enters the
limit cycle region \footnote{In particular, we define $T_\text{out}$ for the as the time, at which the motor frequency crosses
the $\omega_\text{out}=$-1.5 border before entering the limit cycle.}. The machine voltages not shown here
behave analogously.
Before the escape, the system was able to withstand several bursty time intervals.
Thus, escaping is not simply a matter of the actual magnitude of the feed-in noise,
but essentially depends on the system's location in phase space determined by its response to the previous fluctuations
and the actual volume of the stability basin.\\
\indent We calculated the average escape time $\bar T_\text{out}$ as a function of intermittence strength $D$ and 
for Gaussian53 noise for different penetrations $p$. 
As qualitatively expected, the higher the percentage of
fluctuating input, the lower the system stability with respect to noise-induced desynchronization. Additionally,
the system is destabilized with increasing $D$, whereas
for weakly intermittent noise $\bar T_\text{out}$ tends towards the value for Gaussian53 noise. 
For the maximum penetration $p=0.7$, the Gaussian53 scenario overestimates the average escape time of strongly 
intermittent feed-in ($D=2.0$) by a factor of about 1.5, with decreasing penetration this difference grows. 
If we neglect the rated power bound and allow for more extreme events, 
$\bar T_\text{out}$ will decrease.
For Gaussian white noise no escapes are observed for simulations of $T\sim (10^7)$. 
Increasing the standard deviation by a factor of about 5 while dropping the boundedness yields average escape times
of the same order of magnitude as for intermittent and Gaussian53 feed-in.\\
\indent\textit{IEEE grid.}--  
The generator feed-in is $P_{m,M}=0.7$, the motor consumption is set $P_{m,M}=-0.575$ to ensure power
balance, further $B_{ij}=1.3$ for $i\neq j$ and $B_{ij}=0.2-1.3k_i$ for $i=j$ (with $k_i$ being the number of nodes adjacent to 
node $i$).
We now vary the percentage of stochastic feed-in by
the number of fluctuating generators $N_\text{fl}$.
The overall system behaviour in terms of 
phase coherence can be condensed in the order parameter $r(t)=\frac 1N\sum_{i=1}^N\text e^{\ie\delta_i}$ \cite{kurths2003book}.
However, it turnes out reasonable to monitor system
stability based on the individual machines rather than $r(t)$. Thus, we define the \textit{first-escape time} $T_{out}$ 
as the time at which the first machine falls out. Fig.\,\ref{fig:T_out_Nfl} shows an example desynchronization scenario. 
We calculate $\bar T_\text{out}$ on the basis of an ensemble
averaging as follows: First we select $N_\text{fl}$ generators to be fluctuating (with maximum penetration $p=0.7$). 
The initial conditions $\{\delta_i,\omega_i,E_i\}_{i\in\{1,..,74\}}$ correspond to the stationary state, which has to be proven the 
most stable in previous simulations.
In a spatially embedded power grid, spatial correlations of the feed-in play a role. As this is a topic on its own, here we restrict ourselves to the two border cases: independent power 
feed-in $\tilde P_i(t)$ and global noise $\tilde P_i(t)=\tilde P(t)\,\forall\,i$.

Fig.\,\ref{fig:T_out_Nfl} shows the average first-escape time $\bar T_{\text{out}}$ as a function of  
$N_\text{fl}$ for strongly and weakly intermittent and Gaussian53 noise.
With increasing number of fluctuating generators the system gets less stable. 
The differences between strongly intermittent and Gaussian53 noise, as obvious in the two-machine system, are not apparent here.
In this meshed topology, which is typical for transmission grids, each node has $k_i>=2$ neighbors. In contrast to the single G-M system, 
in the complex network the machines have the possibility to compensate rapid feed-in changes via all adjacent machines and reroute 
power flows. 
By performing ensemble averaging
over the locations of the fluctuating generators, we postpone the issue of single node stability depending on the
topological position.
For global noise, the system turnes out to be more stable in this specific case.
Other simulations hint, that this preference depends on the specific topology and generation pattern, in particular the position of
the fluctuating generators. This raises a new issue for follow-up research, as in stochastic systems the stability-topology interplay 
additionally depends on the spatial correlation of the feed-in.

\indent\textit{Frequency and voltage quality.--}
\begin{figure}[t]
\includegraphics[scale=0.46]{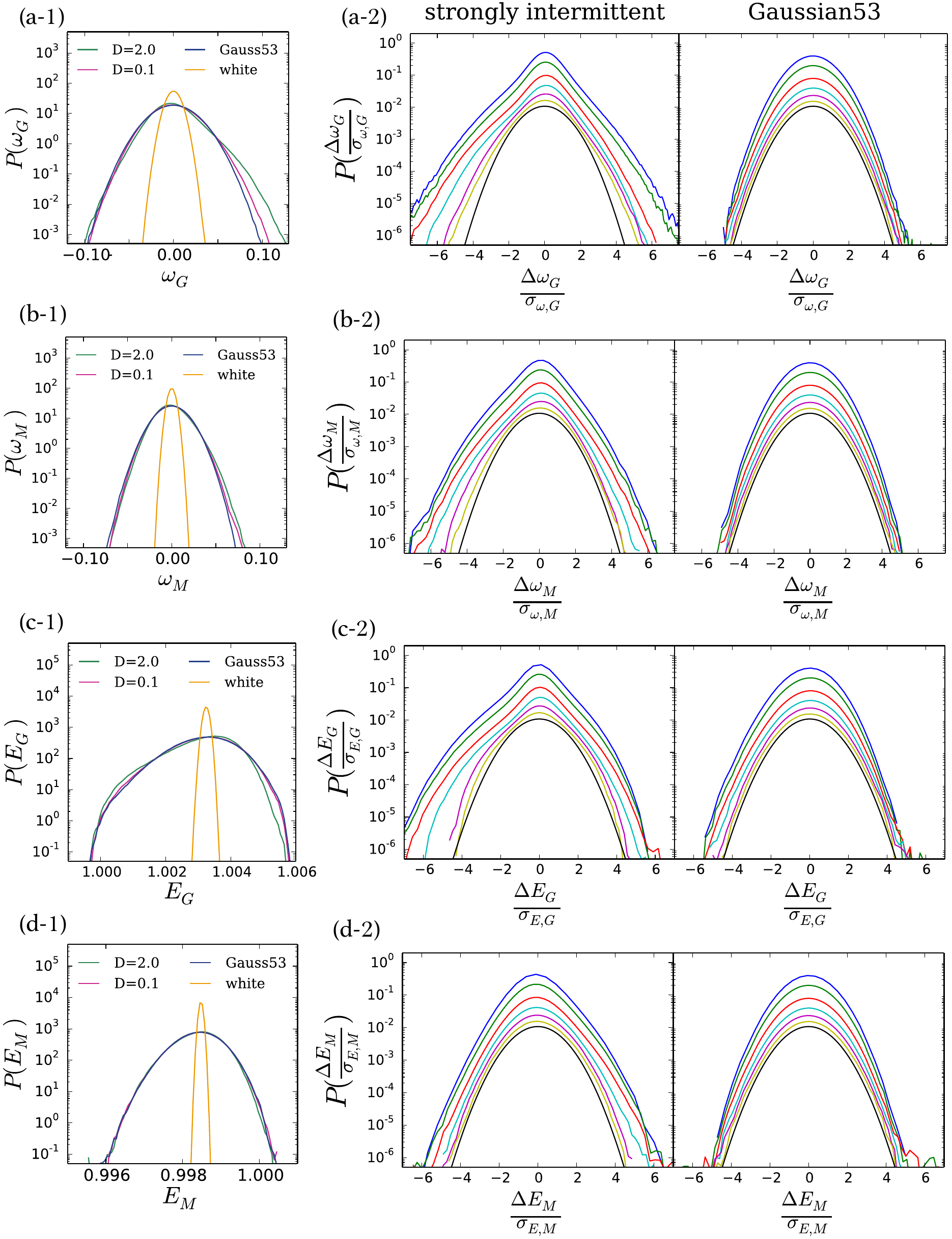}
\caption{Left: Frequency and voltage pdfs for strongly intermittent ($D=2.0$), 
weakly intermittent ($D=0.1$), Gaussian53 and Gaussian white noise.
Right: increment pdfs for $D=2.0$ and Gaussian53 noise with time lags 
$\tau=0.02,0.2,1.0,2.0,5.0,10.0$, Gaussian distribution (top to bottom).}
\label{fig:FreqVoltFluct}
\end{figure}
We select five generators from the IEEE system to be fluctuating and one component consisting of a fluctuating
generator G and one of its adjacent motors M (see Fig.\,\ref{fig:topologies}). We monitor frequency and voltage fluctuations
during normal operation under different noise scenarios with $p=0.2$. 
The frequency and voltage pdfs
give information about the magnitude of the
fluctuations and the probability of violating certain tolerance bounds, the corresponding increment pdfs reveal frequency and voltage
jumps (see Fig.\,\ref{fig:FreqVoltFluct}). 
For all noise scenarios the generator, as the 
primarily affected machine, displays larger fluctuations than the adjacent motor. 
Gaussian white noise underestimates the magnitude of frequency and 
voltage deviations significantly. Gaussian53 noise particularly underrates the probability of large positive frequency deviations.
Even more importantly, the increment pdfs reveal an essential difference between intermittent and non-intermittent 
feed-in: The intermittent
nature of the power fluctuations becomes apparent in the frequency and voltage fluctuations. 
This affects not only the fluctuating generators, but also the other machines in the network. 
Intermittency here implicates, that abrupt large fluctuations occur more likely 
(e.\,g. in Fig.\,\ref{fig:FreqVoltFluct} the probability of an abrupt generator frequency change of $5\sigma_{\omega,G}$
is $\sim 10^3$ larger for strongly intermittent than for Gaussian53 noise). 
This drastically reduces frequency and voltage quality.

\textit{Conclusions.--}
We investigated the impact of intermittent feed-in fluctuations with realistic wind power
properties on grid stability and quality.
With view to system stability, in complex transmission grids severe outages 
due to noise-induced desynchronization are mainly determined by the temporal correlation of the feed-in and the intermittency feature
by itself plays a minor role. However, the intermittent feed-in scenario indicates severe consequences on frequency and voltage 
quality. First,
the likelihood of large frequency deviations, and hence the violation of tolerance bounds, increases. 
But even more importantly,
the intermittent nature of wind is not only transferred into the power output statistics of wind plants \cite{milan2013prl}. 
It also becomes apparent in the increment statistics of frequency and voltage fluctuations.
Electrical engineering communities
are aware of voltage and frequency quality issues due to wind power feed-in, but the substantial implications of intermittency are 
not comprehended \cite{ibrahim2011energyProcedia,albadi2009electricalPowerSystemsResearch}. However, for the design of future energy systems,
it will be absolutely essential to capture intermittency fundamentally.
Our findings also concern solar power input, which is also correlated and even more
intermittent \cite{anvari2016njp}. 
Solar power is fed into the grid via inverters, which lack inertia and therefore aggravate 
stability problems.
In summary, we showed that the characteristics of real wind power destabilize the power grid and decrease frequency and voltage quality.
Further studies, e.\,g. on optimal embedding or smart control and storage, have to take this into account.

\bibliographystyle{unsrt}
\bibliography{bibliothek.bib}

\end{document}